\documentstyle[preprint,prd,aps,epsfig]{revtex}
\begin{document}
\renewcommand{\thesection}{\arabic{section}}
\renewcommand{\thesubsection}{\arabic{subsection}}

\title{ Electrical Conductivity at the Core of a Magnetar}

\author{Sutapa Ghosh$^{a)}$\thanks{E-mail:sutapa@klyuniv.ernet.in},
Sanchayita Ghosh$^{b)}$, Kanupriya Goswami$^{c)}$, 
Somenath Chakrabarty$^{a),d)}$
\thanks{E-mail:somenath@klyuniv.ernet.in} and Ashok
Goyal$^{c),d)}$\thanks{E-mail:agoyal@ducos.ernet.in}\\
a)Department of Physics, University of Kalyani, West Bengal 741 235,
India\\
b)Department of Physics, Ranaghat College, West Bengal, India \\
c)Department of Physics and Astrophysics,
University of Delhi, Delhi 110 007, India \\
d) IUCAA, Post bag 4, Ganeshkhind, Pune 411 007, India}

\maketitle
\begin{abstract}
An expression for the electrical conductivity  at the core of a magnetar is
derived using Boltzmann kinetic equation with the relaxation time 
approximation. The rates for the relevant scattering processes, e.g.,
electron-electron and electron-proton are evaluated in
presence of strong quantizing magnetic fields using tree level diagrams.
It is found that in presence of a strong quantizing magnetic field,
electrical conductivity behaves like  a second rank tensor.
However, if the zeroth Landau levels are only occupied by the charged
particles, it again behaves like a scaler of a one dimensional system.
\end{abstract}

\section{Introduction}
With the recent observational discovery of magnetars, which are assumed
to be strongly magnetized young neutron stars and also the possible
sources of soft gamma-ray repeaters (SGR) and anomalous X-ray pulsars (AXP)
\cite{R1,R2,R3,R4}, the study on the effect of strong
magnetic fields on dense stellar matter has got a new dimension. In the
recent years a lot of work have been done on the effect of strong
magnetic field on the equation of state of dense stellar matter
\cite{R5,R6}. Some
studies have also been done on the effect of strong magnetic field
on elementary processes (e.g., weak and electromagnetic processes) occurring at 
the core region as well as at the envelope of strongly magnetized neutron
stars \cite{R7,R8}. During the past few years, in a number of
publications we have reported our thorough investigations on the effect of 
strong magnetic field on the stability of dense quark matter and quark-hadron 
phase transition (both first order and second order) at the core of a 
strongly magnetized neutron star \cite{R9,R10,R11}. We have also studied
the effect of strong magnetic field on dense neutron matter with properly
modifying the $\sigma - \omega -\rho$-meson type mean-field theory. In those 
work we have developed the relativistic version of mean field theory in
presence of strong quantizing magnetic field using Hartree and Hartree-Fock 
model with $\sigma-\omega-\rho$ mesons exchange interaction
\cite{R12,R13}.

In the recent years the transport coefficients in particular, the electrical 
conductivity of electrons at the crustal region of neutron stars have also 
been calculated in presence of strong magnetic fields \cite{R8}. 
Unfortunately, in none 
of these studies the dense core region of the star is taken into account. Now 
from the observational data of SGR and AXP, the strength of surface magnetic 
field of magnetars are predicted to be $\geq 10^{15}$G. Then it is very easy 
to show by scalar Virial theorem that the magnetic field strength at the core 
region may go up to $10^{18}$G. This is of course strong enough to affect most 
of the physical
processes occurring at the core region. As a consequence, one of the most
important physical phenomena, the transport properties of dense core matter, 
in particular, the electrical conductivity of electrons and transport 
properties of emitted neutrinos will be affected significantly by the strong 
magnetic field. To investigate the transport properties of dense matter
at the core of a neutron star, we use the standard form of Boltzmann kinetic 
equation with relaxation time
approximation. Since protons are too heavy compared to electrons, we
have assumed that the electric current is solely due to the motion of
electrons. Now the relaxation time is directly related to the rates of
the relevant (weak and electromagnetic) processes. these processes are 
strongly affected by the presence of quantizing magnetic field. As a 
consequence, both  qualitative
and the quantitative nature of transport coefficients, e.g.,
electrical conductivity, viscosity and heat conductivity  should change
significantly in presence of strong magnetic field. In a future publication 
we shall report the effect of strong magnetic field on the transport 
properties of neutrinos at the core region \cite{R14}.

In the present article, we shall investigate the effect of strong magnetic
field on the electrical conductivity  at the core region of
strongly magnetized neutron stars or magnetars. The paper is organized in
the following manner: In section 2, we obtain the rates of the relevant
processes in presence of strong magnetic field. In section 3, we shall
derive an expression for the electrical conductivity from Boltzmann
kinetic equation using relaxation time approximation. In the last section,
we have given the conclusion and discussed the future perspective of the work.

\section{Rates}
To obtain electrical conductivity at the core of a magnetar,
we have considered the following two electromagnetic processes relevant
for electron transport: electron-electron scattering
\begin{equation}
e(p_1)+e(p_2) \rightarrow e(k_1)+e(k_2)
\end{equation}
and electron-proton scattering
\begin{equation}
e(p_1)+p(p_2) \rightarrow e(k_1)+p(k_2)
\end{equation} 
where $p_i$ and $k_i$ are the initial and final four momenta. Since the
rate of the electron-neutrino weak scattering, given by
\begin{equation}
e +\nu_e (\bar \nu_e) \rightarrow e +\nu_e (\bar \nu_e)
\end{equation}
is several orders of magnitude less than the electromagnetic processes, it
has almost no significance in the electrical conductivity calculation.
Of course, this is one of the most important neutrino transport processes. 
For the same reason, we have neglected $e-e$ weak scattering.

To obtain electrical conductivity at the core of a magnetar, 
we first consider e-e scattering. For this process,  
both the direct as well as exchange diagrams are considered. From the
definition, the transition matrix element is  given by
\begin{equation}
T_{fi}=-i \int j_\mu^{fi} (x)A^\mu (x) d^4x
\end{equation}
where $j_\mu^{fi}=e \bar \psi_f  \gamma^\mu \psi_i$  is the electric
current produced by the scattered electron and
\begin{equation}
A^\mu (x)=-\int \frac{d^4q}{(2\pi)^4} \frac{\exp[-iq(x-x')]}{q^2} j^\mu
(x')d^4x'
\end{equation}
is the electromagnetic field induced by the current $j^\mu(x')$ of the
other electron and $q$ is the transferred four momentum.

Now in the case, when only the zeroth Landau levels are occupied by the 
electrons the positive energy Dirac spinor is given by
\begin{eqnarray}
\psi^{(+1)}(x) &=& \frac{1}{(L_yL_z)^{1/2}((\varepsilon_0+m
)\varepsilon_0)^{1/2}} \exp[-i(p_0t+p_y y+p_z
z)]\nonumber \\  && \left (\frac{eB}{\pi}\right )^{1/4} \exp 
\left [-\frac{1}{2} eB \left (x-\
\frac{p}{eB}\right )^2\right ]
\left (\begin{array}{c}
\varepsilon_0+m\\ 0\\ p_z\\ 0
\end{array} \right )
\end{eqnarray}

where $L_y$ and $L_z$ are the lengths along $y$ and $z$ directions,
$\varepsilon_0=p_0=(p_z^2+m^2)^{1/2}$ is the modified form of
single particle energy, $p_y$
and $p_z$ are $y$ and $z$ components of electron momentum, $B$ is the
strength of magnetic field, assumed to be along positive $z$-direction
(we have chosen the gauge $A^\mu=(0,0,xB,0)$). The counter part of this
spinor has no contribution when only the zeroth Landau level is occupied. 
Whereas, for low temperature case (the chemical potential $\mu >> T$,
where $T$ is the temperature of the system. In the numerical
calculation, we have taken $T=20$MeV and found that the results are
almost temperature independent for $5{\rm{MeV}}\leq T\leq 30{\rm{MeV}}$) 
the negative energy spinor solutions may be neglected.

Combining eqns.(4)-(6) and after integrating over $t'$, $y'$, $z'$
(which result three $\delta$-functions), we have the transition matrix
element for the direct $e-e$ scattering diagram
\begin{eqnarray}
T_{fi}&=&\-i\int\frac{d^4q d^4x}{2\pi}\delta(q^0-p_1^0+k_1^0)\delta(q_y-p_{1y}
+k_{1y})\delta(q_z-p_{1z}-k_{1z})
\nonumber \\&&\frac{\exp[i q x]}{q^2}\frac{\exp[-i q_x x']}{L_y^2
L_z^2}\frac{e^2}{[16 
p_1^0 p_2^0 k_1^0 k_2^0
(p_1^0+m)(k_1^0+m)(p_2^0+m)(k_2^0+m)]^{1/2}}\nonumber \\
&& \exp[-i\{(p_2^0-k_2^0)t-(p_{2y}-k_{2y})y-(p_{2z}-k_{2z})z\}]
[J_\mu(x')]_{p_1,k_1} [J_\mu(x)]_{p_2,k_2} dx'
\end{eqnarray}
where 
\begin{equation}
[J_\mu(x)]_{p,k}=\bar u(k)\gamma^\mu u(p) I_{0;k_y,k_z}(x) I_{0;p_y,p_z}(x),
\end{equation}
$u$ and $\bar u$ are respectively the positive energy spinor and
the corresponding adjoint and
\begin{equation}
I_{0;k_y,k_z}=\left ( \frac{eB}{\pi} \right )^{1/4} \exp \left [
-\frac{eB}{2} \left ( x-\frac{k_y}{eB}\right )^2 \right ]
\end{equation}
The integrals over $x$ and $x'$ can very easily be evaluated by changing
the variables to $r$ and $R$, given by
$r = (x-x')$ and $R = (x+x')/2$. The integration over $R$ the gives
\begin{equation}
I_R= \left (\frac{\pi}{2 e B}\right )^{1/2}\exp\left [
\frac{1}{2eB}(p_{1y}+p_{2y})^2\right ]
\end{equation}
Similarly, the integration over $r$ is given by
\begin{equation}
I_r=\left (\frac{\pi}{2eB}\right )^{1/2} \exp[-2eBX_{(dir)}^2]
Erfc [-(2eB)^{1/2}X_{(dir)}]
\end{equation}
where $X_{(dir)}=(k_{2y}-p_{1y}-K_{(dir)})/2eB$,  $K_{(dir)}^2=
q_y^2+q_z^2-q_0^2$ and $Erfc(x)$ is the complementary error function.
After evaluating the integrals over $d^4q$ with the help of 
$\delta$-functions we have the transition matrix element for direct 
$e-e$ scattering
\begin{eqnarray}
T_{fi}^{(dir)}&=& -i\frac{e^2}{[16 p_1^0 p_2^0 k_1^0
k_2^0(p_1^0+m)(p_2^0+m)(k_1^0+m)(k_2^0+m)]^{1/2}}\nonumber
\\ && [J_\mu(p_1,k_1)][J_\mu(p_2,k_2)]\frac{1}{4L_y^2L_z^2eB} \nonumber
\\ && \exp\left [\frac{1}{2eB}(p_{1y}+p_{2y})^2\right
]\exp(-2eBX_{(dir)}^2)\nonumber
\\ && Erfc (-(2eB)^{1/2} X_{(dir)})
\frac{1}{2K_{(dir)}}(2\pi)^3
\delta(p_1^0+p_2^0-k_1^0-k_2^0)\nonumber \\
&&\delta(p_{1y}+p_{2y}-k_{1y}-k_{2y})
\delta(p_{1z}+p_{2z}-k_{1z}-k_{2z})
\end{eqnarray}
Which  may be  written as
\begin{equation}
\ T_{fi}^{(dir)}=\Pi^{(dir)} (2\pi)^3
\delta(p_1^0+p_2^0-k_1^0-k_2^0)\delta(p_{1y}+p_{2y}-k_{1y}-k_{2y})\delta(p_{1z}+p_{2z}-k_{1z}-k_{2z})
\end{equation}
The transition matrix element for $e-e$ exchange interaction can very easily
be obtained from the direct one just by exchanging $k_1$ and $k_2$ ($k_1
\leftrightarrow k_2$). Then we have
\begin{eqnarray}
T_{fi}^{(ex)}&=& -i\frac{e^2}{[ 16 p_1^0 p_2^0 k_1^0
k_2^0(p_1^0+m)(p_2^0+m)(k_1^0+m)(k_2^0+m)]^{1/2}}\nonumber
\\ && [J_\mu(p_1,k_2)][J_2(p_2,k_1)]\frac{1}{4 L_y^2 L_z^2 eB}\nonumber
\\ &&\exp\left [\frac{1}{2eB}(p_{1y}+p_{2y})^2\right ]
\exp(-2eBX_{(ex)}^2)
Erfc[-(2eB)^{1/2}X_{ex)}]\frac{1}{2K_{(ex)}}
(2\pi)^3 \nonumber \\ &&
\delta(p_1^0+p_2^0-k_1^0-k_2^0)\delta(p_{1y}+p_{2y}-k_{1y}-k_{2y})
\delta(p_{1z} +p_{2z}-k_{1z}-k_{2z})
\end{eqnarray}
where $X_{(ex)}=(k_{1y}-p_{1y}-K_{(ex)})/2eB$ and $K_{(ex)}=
K_{(dir)}$. Which may also be  written in the 
form
\begin{eqnarray}
T_{fi}^{(ex)}&=&\Pi^{(ex)}
(2\pi)^3 \delta(p_1^0+p_2^0-k_1^0-k_2^0)\delta(p_{1y}+p_{2y}-k_{1y}-k_{2y})
\nonumber \\ &&\delta(p_{1z} +p_{2z}-k_{1z}-k_{2z})
\end{eqnarray}
Now from the elementary  definition of rate per unit volume for the 
scattering process, we have the 
differential rate for $e-e$ scattering
\begin{eqnarray}
dW&=&\left | \Pi^{(dir)}+\Pi^{(ex)}\right |^2
(2\pi)^3 \delta(p_1^0+p_2^0-k_1^0-k_2^0)\delta(p_{1y}+p_{2y}-k_{1y}-k_{2y})
\delta(p_{1z} +p_{2z}-k_{1z}-k_{2z}) \nonumber \\
&&\frac{dk_{1y} dk_{1z} dk_{2y} dk_{2z} dp_{2y} dp_{2z}}{(2\pi)^6}
\end{eqnarray}
Hence to obtain the differential rate $dW$ of the processes we have to 
calculate $\mid \Pi^{(dir)}\mid^2$, $\mid \Pi^{(ex)} \mid^2$ and the cross term.
Further, to compute these quantities, we have to evaluate the traces 
of the products of $\gamma$-matrices.
Using the standard techniques, we have computed the traces of the products of  
$\gamma$-matrices.  Then we have
\begin{eqnarray}
dW_{(dir)}&=&\frac{e^4}{p_1^0p_2^0k_1^0k_2^0} \exp[\frac{1}{eB} (p_{1y}
+p_{2y})^2 ] \exp(-4eBX_{(dir)}^2) \nonumber \\
&& \{Erfc[-(2eB)^{1/2} X_{(dir)}]\}^2
[(p_1.p_2)(k_1.k_2)+(p_1.k_2)(k_1.p_2)-m^2(p_2.k_2)-m^2(p_1,k_1)
\nonumber\\ &&
+2m^4] \frac{1}{8 \pi^2 K_{(dir)}^2}  \frac{1}{(2\pi)^3}
\delta(p_1^0+p_2^0-k_1^0-k_2^0)\delta(p_{1y}+p_{2y}-k_{1y}-k_{2y})
\nonumber\\ &&\delta(p_{1z} +p_{2z}-k_{1z}-k_{2z}) dk_{1y} dk_{1z} 
dk_{2y} dk_{2z} dp_{2y} dp_{2z}
\end{eqnarray}

Similarly, the exchange term is obtained by $k_1  \leftrightarrow  k_2$
and is given by
\begin{eqnarray}
dW_{(ex)}&=&\frac{e^4}{p_1^0 p_2^0 k_1^0
k_2^0}\exp[\frac{1}{eB}(p_{1y}+p_{2y})^2]\exp(-4eBX_{(ex)}^2)
{Erfc[-(2eB)^{1/2}X_{(ex)}]}^2 \nonumber \\ &&
[(p_1.p_2)(k_1.k_2)+(p_1.k_1)(p_2.k_2)-m^2(p_2.k_1)-m^2
(p_1.k_2)+2m^4]\frac{1}{8\pi^2
K_{(ex)}^2}\nonumber \\ && 
\frac{1}{(2\pi)^3}
 \delta(p_1^0+p_2^0-k_1^0-k_2^0)\delta(p_{1y}+p_{2y}-k_{1y}-k_{2y})
\delta(p_{1z} +p_{2z}-k_{1z}-k_{2z})\nonumber \\ &&
dk_{1y} dk_{1z} dk_{2y} dk_{2z} dp_{2y} dp_{2z}
\end{eqnarray}

Finally the cross term whose evaluation is a bit lengthy, but 
straight-forward, is given by
\begin{eqnarray}
dW_{(cross)}&=&\frac{e^4}{p_1^0 p_2^0 k_1^0
k_2^0}\exp\left [\frac{1}{eB}(p_{1y}+p_{2y})^2\right ]
\exp[-2eB(X_{(dir)}^2 +X_{(ex)}^2)]\nonumber  \\
&& Erfc[-(2eB)^{1/2}X_{(dir)}]Erfc[-(2eB)^{1/2}X_{(ex)}]\nonumber \\&&
[-2(k_1.k_2)(p_1.p_2)+m^2(p_2.k_2)+m^2(p_2.k_1)+m^2(p_2.p_1)+\nonumber
\\ && m^2(p_1.k_1)+m^2(p_1.k_2)+m^2(k_1.k_2)-2m^4] \nonumber \\
&&\frac{1}{8\pi^2 K_{(dir)}K_{(ex)}}\frac{1}{(2\pi^3)}
\delta(p_1^0+p_2^0-k_1^0-k_2^0)\delta(p_{1y}+p_{2y}-k_{1y}-k_{2y})
\delta(p_{1z} +p_{2z}-k_{1z}-k_{2z})\nonumber \\ &&
dk_{1y} dk_{1z} dk_{2y} dk_{2z} dp_{2y} dp_{2z}
\end{eqnarray}

The simplified form of these rates can very easily be obtained by integrating 
over the momentum components $k_{1z}$, $k_{2z}$, $p_{2y}$ and $p_{2z}$ with the help of $\delta$-functions. Then we have after some
simple algebraic manipulation
\begin{eqnarray}
dW_{(dir)}&=& \frac{e^4}{p_1^0k_1^0p_2^0k_2^0}
\exp\left [\frac{1}{eB}(k_{1y}+k_{2y})^2 \right ]
\exp\left [\frac{1}{eB}(k_{2y}-k_{1y}-K_{(dir)})\right ]\nonumber  \\
&& \left [Erfc\left \{-\frac{1}{(2eB)^{1/2}}(k_{2y}-k_{1y}-K_{(dir)})\right
\}\right]^2
[(p_1^0p_2^0-p_{1z}p_{2z})(k_1^0k_2^0-k_{1z}k_{2z})+ \nonumber \\
&&(p_1^0k_2^0-p_{1z}k_{2z})(k_1^0p_2^0-k_{1z}p_{2z})
-m^2(p_2^0k_2^0-p_{2z}k_{2z}) -m^2(p_1^0k_1^0 -p_{1z}k_{1z}) +2m^4]
\nonumber \\
&& \frac{1}{8\pi^2K_{(dir)}^2}\frac{1}{(2\pi)^3}dk_{1y}dk_{2y}\Huge |_{k_{1z}
=p_F, k_{2z}=k_{2z}^{(R)}, p_{2z}=p_{1z}+k_{1z}-k_{2z}}\nonumber \\
&& \frac{p_F}{\mid f'(k_{2z})\mid_{k_{2z}=k_{2z}^{(R)}}}
\end{eqnarray}
where $K_{(dir)}^2=(p_{1z}-p_F)^2-(p_1^0-\mu)^2$, $\mu$ and $p_F$ are
the electron chemical potential and Fermi momentum respectively,
\begin{equation}
k_{2z}^{(R)}=\frac{m(p_1^0+\mu)}{[2m^2+2p_1^0\mu -2p_{1z} p_F]^{1/2}}
\end{equation}
is the root of the transcendental equation $p_1^0+p_2^0-k_1^0-k_2^0=0$ and
\begin{equation}
\left | f'(k_{2z})\right |_{k_{2z}=k_{2z}^{(R)}}=\left | \frac{(b-k_{2z}^{(R)})}
{\{(b-k_{2z}^{(R)})^2+m^2\}^{1/2}} -\frac{k_{2z}^{(R)}} {(k_{2z}^{(R)2}
+m^2)^{1/2}} \right |
\end{equation}
where $b=p_{1z}+p_F$

Similarly, for the exchange interaction we have
\begin{eqnarray}
dW_{(ex)}&=&\frac{e^4}{p_1^0k_1^0p_2^0k_2^0}\exp\left [ \frac{1}{eB}
(k_{1y}+k_{2y})^2\right ] \exp\left [
-\frac{1}{eB}(k_{1y}-k_{2y}-K_{(ex)})^2 \right ]\nonumber \\
&& \left \{Erfc\left [ -\frac{k_{1y}-k_{2y}-K_{(ex)}}{(2eB)^{1/2}} \right ]
\right \}^2
[(p_1.p_2)(k_1.k_2) +(p_1.k_1)(p_2.k_2)\nonumber \\
&& -m^2 (p_2.k_1)-m^2(p_1.k_2) +2m^4] \frac{1}{8\pi^2 K_{(ex)}^2}
\frac{1}{(2\pi)^3} \nonumber \\
&& dk_{1y}dk_{2y}\Huge |_{k_{2z}=k_F, k_{1z}=k_{1z}^{(R)}, k_{2z}=
k_{1z}+ p_{1z} -k_{2z}} \frac{p_F}{\left | f'(k_{2z})\right |_{k_{2z} =
k_{2z}^{(R)}} }
\end{eqnarray}
where $K_{(ex)}^2=(p_{1z}-p_F)^2- (p_1^0-\mu)^2$. Finally for the mixed
term, we have
\begin{eqnarray}
dW_{(cross)}&=&\frac{e^4}{p_1^0k_1^0p_2^0k_2^0}\exp\left [ \frac{1}{eB}
(k_{1y}+k_{2y})^2\right ] 
\exp\left [ -\frac{1}{2eB}(k_{2y}-k_{1y}-K_{(dir)})^2 \right ] \nonumber
\\
&&\exp\left [ -\frac{1}{2eB}(k_{1y}-k_{2y}-K_{(ex)})^2 \right ]
Erfc\left [ -\frac{k_{2y}-k_{1y}-K_{(dir)}}{(2eB)^{1/2}} \right ]
\nonumber \\ &&
Erfc\left [ -\frac{k_{1y}-k_{2y}-K_{(ex)}}{(2eB)^{1/2}} \right ]
[-2(p_1.p_2)(k_1.k_2) +m^2(p_2.k_2)
 +m^2 (p_2.k_1)+\nonumber \\ &&m^2(p_1.p_2)+m^2(p_1.k_1) +m^2(p_1.k_2)
+m^2(k_1.k_2)-2m^4] \nonumber \\ &&\frac{1}{8\pi^2 K_{(dir)}K_{(ex)}}
\frac{1}{(2\pi)^3} \nonumber \\
&& dk_{1y}dk_{2y}\Huge |_{k_{2z}=k_F, k_{1z}=k_{1z}^{(R)}, k_{2z}=
k_{1z}+ p_{1z} -k_{2z}} \frac{p_F}{\left | f'(k_{1z})\right |_{k_{1z} =
k_{1z}^{(R)}} }
\end{eqnarray}
In the case of $e-p$ scattering, given by eqn.(2)
the direct diagram only contributes and the form of  trace term will be
slightly different
from that of $e-e$ direct term,  given by
\begin{eqnarray}
& & 32(p_1^0+m)(k_1^0+m)(p_2^0+M^*)(k_2^0+M^*)[(p_1.p_2)(k_1.k_2)+\\
& & (p_1.k_2)(k_1.p_2)-m^2(p_2,k_2)-M^{*2}(p_1.k_1)+2m^2 M^{*2}]
\end{eqnarray}
where $M^*$ is the effective proton mass in the Hartree type mean field
model in presence of strong magnetic field \cite{R12}. In this case we have 
to use $p_2^0= (p_{2z}^2+M^2)^{1/2}$ and $k_2^0= (k_{2z}^2+M^2)^{1/2}$ and make
necessary changes in eqn.(20) of $dW_{(dir)}$ for $e-e$ scattering to
obtain $dW_{(dir)}^{(ep)}$. In the next section we shall use these
differential rates (eqns.(20), (23), (24) and $dW_{(dir)}^{(ep)}$)
to obtain the electrical conductivity.

\section{Electrical Conductivity}
To compute  the electrical conductivity of the magnetized stellar
matter, we  consider the
simplest form of Boltzmann kinetic equation
for  electrons which are assumed to be slightly out of local
thermodynamic  equilibrium. The kinetic
equation is then given by \cite{R15,R16}
\begin{equation}
\frac{\partial f}{\partial t} + \frac{p_i}{p_0} \nabla_i f+
\dot p_i\frac{\partial f }{\partial p_i }=C
\end{equation}
Where $f$ is the non-equilibrium distribution function and $C$ is the
collision term, given by
\begin{eqnarray}
C&=&\int \frac{d^3p_2}{(2\pi)^3} \frac{d^3p_3}{(2\pi)^3} \frac{d^3p_4}{(2\pi)^3}
(2\pi)^3\delta(p_1^0+p_2^0-k_1^0-k_2^0)\nonumber \\ && 
\delta(p_{1y}+p_{2y}-k_{1y}-k_{2y}) \delta(p_{1z}+p_{2z}-k_{1z}-k_{2z})
\nonumber \\ && \mid T_{fi}\mid^2 {\cal{F}}
\end{eqnarray}
where ${\cal{F}}= [f_3f_4(1-f_1)(1-f_2) -f_1f_2(1-f_3)(1-f_4)]$, and
$(1-f_i)$'s are the Pauli blocking factors.
Replacing collision term by the  relaxation time approximation, we have the 
relevant portion of the kinetic equation needed to obtain the electrical
conductivity.
\begin{equation}
\dot p_i\frac{\partial f }{\partial p_i }=-\frac{(f-f_0)}{\tau}
\end{equation}
where $f_0$ is the local equilibrium distribution function for electrons
(Fermi distribution), given by
\begin{equation}
f_0=\frac{1}{\exp(\beta(x) ((p_z^2+p_\perp^2+m^2)^{1/2}-\mu(x) ))+1}
\end{equation}
Here $p_z$ is the longitudinal momentum and $p_\perp=(2\nu
eB)^{(1/2)}$ is the corresponding transverse part for the
electrons, $T(x)$=$1/\beta(x)$ and $\mu(x)$ are the local temperature
and chemical potential.
When only  the zeroth Landau levels ($\nu=0$) are occupied by the electrons, 
the transverse momentum becomes exactly zero.

We now consider the first order approximation, i.e., the system is very close to
its local equilibrium configuration, then we can write
$f(p,x)=f_0(p)+\delta f(p,x)$ 
where $\delta f$ is a measure of its deviation from local statistical 
equilibrium configuration. Then we have
${\cal{F}}= -(\delta f)\{f_{3,0}f_{4,0}(1-f_{2,0}) -f_{2,0}(1-f_{3,0})
(1-f_{4,0})\}$, where $f_{i,0}$'s are the equilibrium distribution
function for the $i$th component (Fermi distribution). The we can write
\begin{equation}
\frac{\partial f_0 }{\partial p_i }=-f_0(1-f_0)\beta\frac{p_i}{\varepsilon}
\end{equation}
Expressing the force term in the form, $\dot p_i=eE_i$ we have after
using eqn.(3)
\begin{equation}
\frac{\delta f }{\tau}=f_0(1-f_0)\beta qE_i\frac{p_i}{\varepsilon}
\end{equation}
Then just by inspection, it is very easy to realize that the relaxation time is
obtained by evaluating the integrals over $dk_{1y}$ and $dk_{2y}$  in
the expressions for the rates and the relation with the total rate of the 
processes is given by
\begin{equation}
\tau(p_{1z})= \frac{1}{W}
\end{equation}
where $W=W_{(dir)}^{(ee)} +W_{(ex)}^{(ee)} +W_{(cross)}^{(ee)}
+W_{(dir)}^{(ep)} $
therefore,
\begin{equation}
\frac{1}{\tau} =\frac{1}{\tau_{(dir)}^{(ee)}}
+\frac{1}{\tau_{(ex)}^{(ee)}}
+\frac{1}{\tau_{(mix)}^{(ee)}} +\frac{1}{\tau_{(dir)}^{(ep)}}
\end{equation}
the well know Matthiessen's  rule \cite{R17}. Now the expression for 
electric current is given by
\begin{equation}
j_i=e\int f(x,p) \frac{p_i}{\varepsilon}\frac{d^3p}{(2\pi)^3}
\end{equation}
Because of randomness in the system, the local current vanishes at equilibrium. 
Then the nonzero part of electric current is obtained by using the first order 
deviation $\delta f(p,x)$ and is given by 
\begin{equation}
j_i=\frac{e^2\beta}{(2\pi)^3}\int \frac{p_i}{\varepsilon}\tau
f_0(1-f_0)\frac{p_j}{\varepsilon} d^3p E_j
\end{equation}
In presence of a strong magnetic field of strength $B > B_c^{(e)}$, the
quantum limit of magnetic field, where $B_c^{(e)}=4.4\times 10^{13}$G, we 
have $d^3p=2\pi eB dp_z$. Then writing  Ohm's law in the form $j_i=\sigma_{ij} 
E_j$, we have
\begin{equation}
\sigma_{ij}=\frac{e^2\beta eB}{2\pi^2}\int \tau
f_0(1-f_0)\frac{p_i p_j}{\varepsilon^2} dp_z
\end{equation}
Which is second rank tensor with the components $\sigma_{zz}$,
$\sigma_{z\perp}$, $\sigma_{\perp z}$ and $\sigma_{\perp \perp}$.
Since we are interested for $\nu = 0$ case only, i.e., when the lowest Landau 
levels are populated, we have $\sigma_{\perp \perp}=\sigma_{\perp z}=
\sigma_{z\perp}=0$ and
\begin{equation}
\sigma_{zz}=\frac{e^2\beta}{2\pi^2}eB\int \tau(p_z)f_0(1-f_0)
\frac{p_z^2}{ \varepsilon^2}dp_z
\end{equation}
which is the only non-zero component. The electron transport in this
special case becomes essentially one dimensional- along the direction of
magnetic field.

In fig.(1) we have plotted the variation of $\sigma_{zz}$ with electron 
density for (a)$B=10^3\times B_c^{(e)}$, (b)$B=5\times 10^3\times B_c^{(e)}$,
(c)$B=10^4\times B_c^{(e)}$, (d)$B=5\times 10^4\times B_c^{(e)}$ and (e) 
$B=10^5\times B_c^{(e)}$, where $B_c^{(e)}\approx 4.4\times 10^{13}$G,
the quantum limit for electrons. In fig.(2) we have shown the variation of 
the nonzero component of electrical conductivity $\sigma_{zz}$ with magnetic 
field strength for (a)$n_e=n_e^0$, (b)$n_e=10n_e^0$, (c)$ne=50n_e^0$, and 
(d)$n_e=100n_e^0$, where $n_e^0=10^{-2}$fm$^{-3}$, the typical electron 
density in a non-magnetic neutron star. In both these cases we have not taken
$\beta$ equilibrium condition into account.  The density of electron is in 
some sense arbitrary. Therefore, we now consider an interacting
$n-p-e$ system in $\beta$-equilibrium, which is the real physical picture 
at the core region of a neutron star. Then we have $n_p=n_e$, the charge
neutrality condition, $\mu_n=\mu_p+\mu_e$, the $\beta$-equilibrium
condition (we assume that the neutrinos are non-degenerate, leave the system 
as soon as they are produced) and baryon number density $n_B=n_p+n_n$ remains 
invariant in electromagnetic and weak interaction processes. Solving these 
constraints self-consistently at the core of a magnetar, we obtain the 
chemical potential of the constituents and hence the density of electrons for 
various baryon densities and magnetic field strengths. We have assumed a 
fixed temperature $T=20$MeV. In fig.(3) we have shown the variation of 
$\sigma_{zz}$ with the strength of magnetic field for a fixed baryon number 
density (in the numerical computation we have taken $n_B=4n_0$, $n_0=
0.17$fm$^{-3}$) assuming that the system is in $\beta$-equilibrium.
In fig.(4) we have plotted $\sigma_{zz}$ against the electron density
for a fixed magnetic field strength ($B=10^5B_c^{(e)}$) in $\beta$-equilibrium 
condition.

\section{Conclusion}
We have studied the electrical conductivity at the core of a magnetar
using Boltzmann kinetic equation with the relaxation time approximation. The
latter is obtained from the rates of the electromagnetic processes, which 
essentially control the electron transport in the medium. We have
further noticed that the electrical conductivity behaves like a second
rank tensor above the critical value (quantum limit) of magnetic field
strength. However, only $\sigma_{zz}$ components is non-zero when the
lowest Landau levels are occupied by the charged particles. The system
effectively becomes one-dimensional in presence of quantizing magnetic
field- electrons can move along the field direction. The electric
current vanishes in the plane transverse to the direction of magnetic
field. Since it is impossible to compute the rates analytically (even it 
is very difficult to obtain numerically) with the non-zero values of Landau 
quantum numbers for all the four charged particles, we have assumed that only 
the zeroth Landau levels are occupied. This makes our like much simpler. The
assumption of only zero Landau quantum number is also justified  by the
presence of intense magnetic field at the core region of magnetars.
Fig.(1) shows that there exist cut off densities, beyond which, the
Pauli blocking factor supresses the electron scattering and as a result
the conductivity becomes zero. This figure also shows that the cut off
density increases with the increas of magnetic field strength. As the
magnetic field increases, for a given baryon number density, the
chemical potential of electrons decreases, which further reduces the
Pauli suppression factor. Similarly, we have noticed from fig.(2) that
there is a minimum value for magnetic field strength for a particular
baryon number density below which the conductivity again becomes zero
because of same reason as discussed above. On the other hand, in the
case of $\beta$-equilibrium condition, electrons are generated
self-consistently. So there are no such cut off values for the density or
magnetic field and the variation is smooth. 

With this simplified picture, we have obtained the rates and finally
$\sigma_{zz}$ for various baryon densities and magnetic field strengths.

\begin{figure} 
\psfig{figure=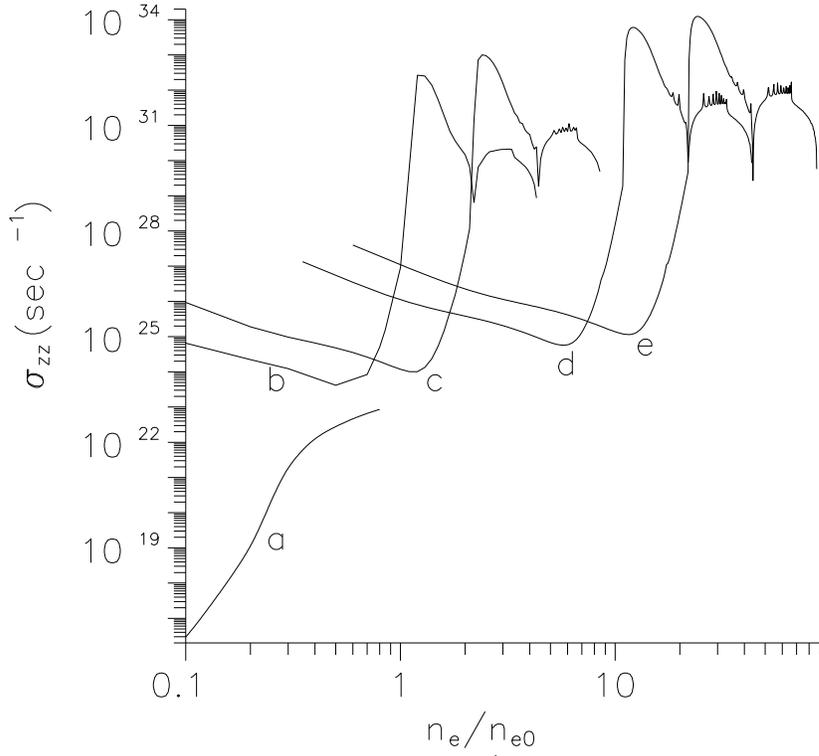,height=0.6\linewidth}
\caption[]{Variation of $\sigma_{zz}$ in sec$^{-1}$ with electron density
for (a)$B=10^3\times B_c^{(e)}$, (b)$B=5\times 10^3\times B_c^{(e)}$,
(c)$B=10^4\times B_c^{(e)}$, (d)$B=5\times 10^4\times B_c^{(e)}$  and
(e)$B=10^5\times B_c^{(e)}$.}
\end{figure}
\begin{figure} 
\psfig{figure=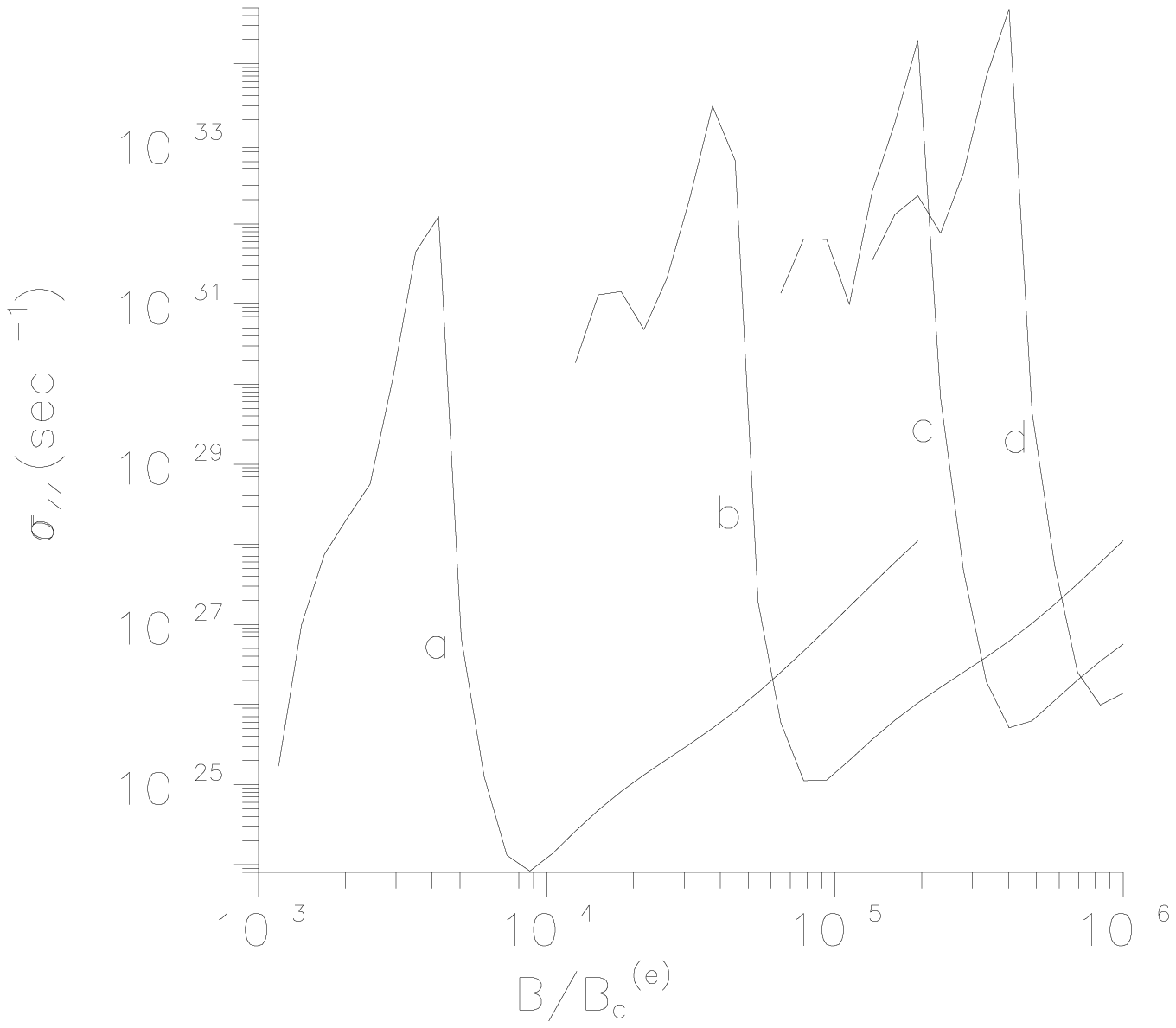,height=0.6\linewidth}
\caption[]{Variation of$\sigma_{zz}$ in sec$^{-1}$ with magnetic field
strength,
for (a)$n_e=n_e^0$, (b)$n_e=10n_e^0$,(c)$n_e=50n_e^0$ and (d)$n_e=100n_e^0$.}
\end{figure}
\begin{figure} 
\psfig{figure=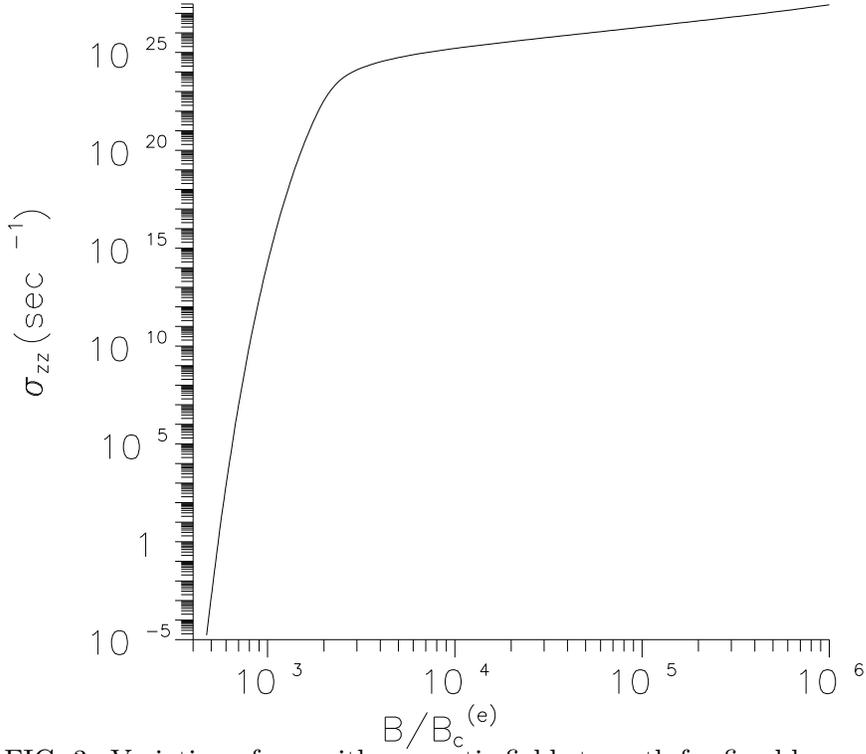,height=0.6\linewidth}
\caption[]{Variation of $\sigma_{zz}$ with magnetic field strength for
fixed baryon density ($n_B=4n_0$). The system is assumed to be in
$\beta$-equilibrium.}
\end{figure}
\begin{figure} 
\psfig{figure=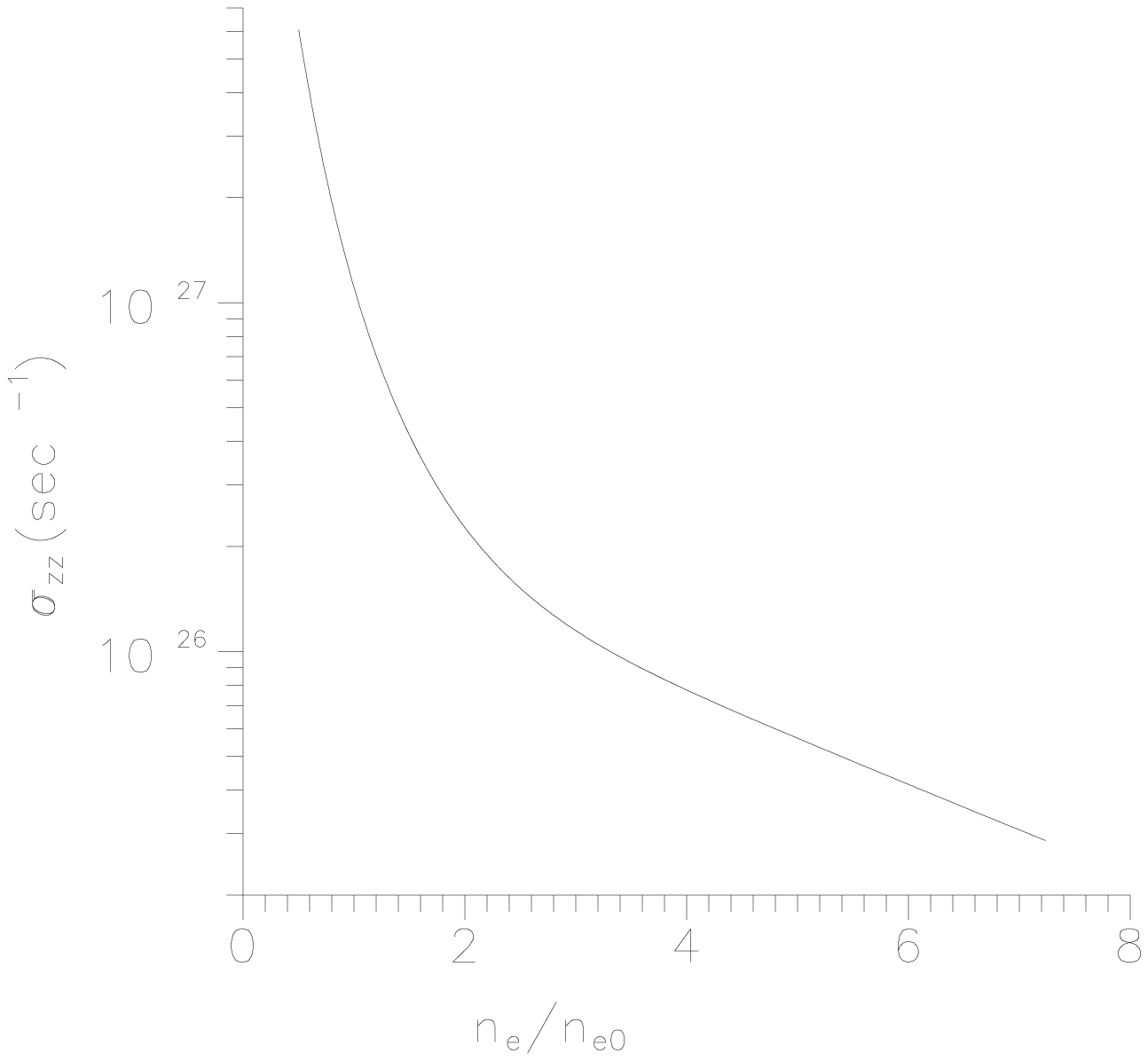,height=0.6\linewidth}
\caption[]{Variation of $\sigma_{zz}$ with electron density for constant
magnetic field strength ($B=10^5\times B_c^{(e)})$.  The system is assumed to 
be in $\beta$-equilibrium.} 
\end{figure}

\noindent Acknowledgment: SC is thankful to Department of Science and 
Technology, Govt. of India, for partial support of this work, Sanction 
number:SP/S2/K3/97(PRU).  
\end{document}